\documentclass{optica-article}
\journal{boe}
\articletype{Research Article}
\usepackage{graphicx,caption}
\usepackage{enumitem}
\begin{document}
\title{Polarised light modular microscopy (Pol-ModMicro) for identifying hemozoin crystals in \textit{Plasmodium}}

\author{Fraser Eadie,\authormark{1} Matthew P. Gibbins,\authormark{2, 3} Graham M. Gibson,\authormark{1} Matthias Marti,\authormark{2,3} and Akhil Kallepalli\authormark{1,*}}

\address{\authormark{1} School of Physics and Astronomy, University of Glasgow, Glasgow, G12 8QQ, United Kingdom \\
\authormark{2} Wellcome Centre for Integrative Parasitology, School of Infection and Immunity, University of Glasgow, Glasgow, G12 8TA, United Kingdom \\
\authormark{3} Institute of Parasitology, University of Zurich, Zurich, CH-8057, Switzerland}

\email{\authormark{*} Akhil.Kallepalli@glasgow.ac.uk}
\homepage{https://kallepallilab.com/} 

\begin{abstract*}
\textit{Plasmodium} spp. are the protozoan parasites responsible for malaria. \textit{Plasmodium} spp. synthesise a biocrystal, hemozoin, which can be observed under cross-polarised light. These birefringent crystals can be seen due to different refractive indices of the hemozoin crystal and red blood cells. Here, we present a polarised light modular microscopy (Pol-ModMicro) solution, complete with illumination sources and a robust imaging system to capture birefringence and identify the parasites. We achieve this by combining ModLight light sources with bespoke components designed for the OpenFlexure microscope to image blood smears. Further, a simple and robust algorithm capable of enhancing birefringence is presented. This solution provides image quality that is comparable to a substantially more expensive proprietary microscope. 
\end{abstract*}

\section{Introduction}
\noindent Malaria is one of the most important public health challenges, causing more than 600,000 deaths annually and responsible for 250 million infections worldwide \cite{WHO}. \textit{Plasmodium} spp., the aetiological agent of malaria disease, are Apicomplexan protozoal parasites with a complex life cycle involving both human (or other vertebrates) and mosquito host \cite{Venugopal2020-lm}. The parasite in the blood stages of its life cycle invades and multiplies within red blood cells before rupturing the host cell to repeat this cycle \cite{Venugopal2020-lm}. Within the red blood cell, the parasite digests the abundant protein haemoglobin, which is used to carry oxygen around the body. The parasites require iron and they use this from the haem moiety in haemoglobin. However, haem is inherently toxic as it can generate reactive oxygen species. To overcome this problem, the parasite detoxifies haem to generate hemozoin, packaging this waste product in a crystalline structure \cite{Coronado}. Interestingly, hemozoin is birefringent, and cross-polarised light microscopy has shown promise when used to identify \textit{Plasmodium} parasites \cite{Olson, Gordon}.

Polarised light microscopy allows a unique insight into the analysis of natural and man-made materials \cite{Mehta}, exploiting birefringence which is a result of anisotropic patterns in the atomic structure of a material. This phenomenon occurs due to double refraction of light in transparent, non-isotropic and molecularly ordered materials which create a variation in the intensity of the separate rays of light \cite{Koike}. This difference in intensity is caused by the existence of orientation-dependent refractive indices ($n_1$, $n_2$) of translucent organic and inorganic structures \cite{Douglas}. The intensity difference ($\Delta{I}$) is a linear relationship for the blue and red wavelengths of light \cite{Ross}. 
\begin{equation}
    \textnormal{Birefringence} = n_1 - n_2
\end{equation}

\begin{equation}
    \Delta{I_b} = I_{o_b}[3.6*10^{-3}R_b]
\end{equation}

\begin{equation}
    \Delta{I_r} = I_{o_r}[2.2*10^{-3}R_r]
\end{equation}

\begin{equation}
  R_{b,r} = d(\sqrt{\epsilon_1}-\sqrt{\epsilon_2})    
\end{equation}
\noindent Where $\epsilon_1$ and $\epsilon_2$ are dielectric constants perpendicular and parallel to the optical axis of the biological sample. $R_{b,r}$ are the retardations of the sample for blue and red wavelengths of light, respectively. The birefringence shine is proportional to the difference in optical path length (defined as the distance light needs to travel in the air to have the same phase difference as that of the medium in consideration) and is represented as the distance between the ordinary and extra-ordinary rays, divided by the thickness of the molecular structure ($\Delta t$) \cite{Douglas}. 

\begin{equation}
    \textnormal{Birefringence} \propto OPD = \frac{n_1 - n_2}{\Delta t} 
\end{equation}

The interaction of light within the molecular structure strongly depends upon the direction of the light's electric field ($E$-field). For perfect birefringence to occur the $E$-field must interact exactly perpendicular to the molecular structure of the sample (which may not always be oriented horizontally on the slide when considering its local position). Light travelling with the $E$-field along the perpendicular axis has different principal refractive indices caused by the varying molecular structure of the sample, this difference allows birefringence to be observed. By knowing the optical properties, birefringence can be used in biological samples to analyse specific molecular structures \cite{DOUPLIK201347}. The polarised light creates a much larger emission intensity from the targeted sites, compared with unpolarised light, drastically increasing the imaging clarity of the sample \cite{ROMAGOSA}.

Birefringence can be observed by placing a sample between two orthogonally polarised films, in the simplest case. Practically, this is done by fixing one polariser and rotating the other on opposite sides of the sample to establish cross-polarisation.  In experiments with microscope stage control, volumetric image reconstruction can be done by collecting a Z-stack of images in cross-polarised light at multiple depths. In the case of \textit{Plasmodium}-infected red blood cells, the cells will be highlighted as they `shine' due to birefringence \cite{Douglas, Maude}.

The primary component of this research is the microscope, adopted from the OpenFlexure project \cite{McDermott}. The OpenFlexure 3D printed microscopes range from a basic, manually operated microscope to a motorised version \cite{Robert} and those capable of epifluorescence microscopy. Due to its simplistic design, the OpenFlexure microscope is 3D printable, making it more accessible than a standard microscope costing comparatively higher \cite{McDermott}. In addition, the OpenFlexure microscope makes high-quality imaging \cite{Mullen} accessible, and is portable and easy to maintain \cite{McDermott}. The custom cross-polarising microscope and software are invaluable tools when using the microscope to capture images of several cells and diseases such as malaria \cite{Knapper}. Herein lies the motivation for this study and the solution offered for low-cost identification and imaging of hemozoin in malarial parasites. 


\section{Methods}
The methodology of this study is clearly divided into two segments: (1) adoption and modification of the OpenFlexure microscope for polarised light microscopy, and (2) imaging of hemozoin crystals and identification of malaria parasites distribution. 

\subsection{Microscope Components}
\begin{figure}[t]
\centering
\captionsetup{width=\linewidth}
\includegraphics[width=0.95\columnwidth]{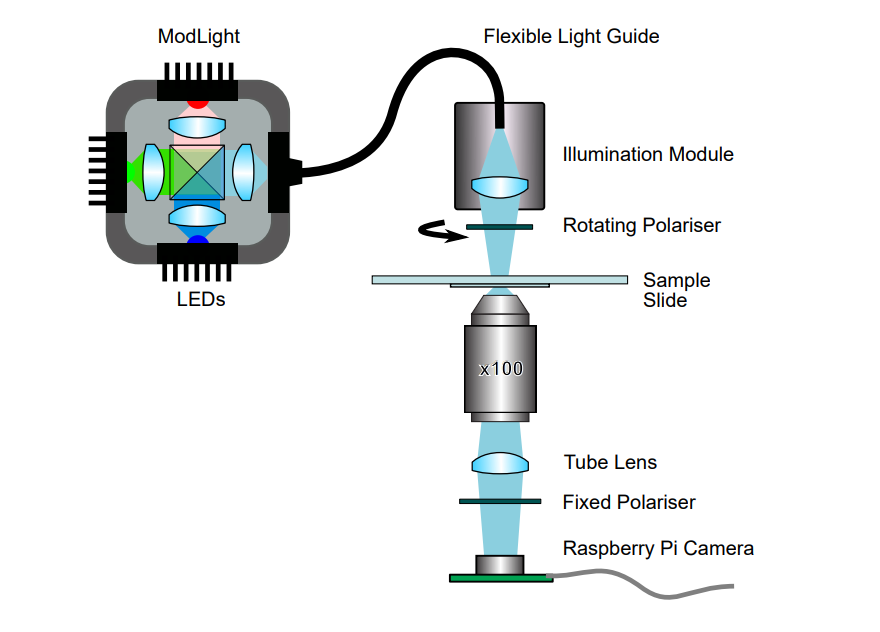}
\caption{\label{Figure 1.} The polarised light modular microscope (Pol-ModMicro) system combines modular light sources (ModLight) \cite{modlight} with a modified OpenFlexure Delta Stage microscope to image the hemozoin crystals. The microscope includes a modified condenser arm to hold the flexible light guide and the rotating polariser and a modified cube for holding the fixed polariser. The Raspberry Pi camera module captures the images and subsequent processing identifies and correlates the polarised light images with the brightfield images for identification and localisation of hemozoin crystals. All the modified components and the OpenFlexure Delta Stage files are available in \cite{Eadie2022}}
\end{figure}

The OpenFlexure platform \cite{Stirling2020,Cicuta2020,Sharkey2016} provides a stable hardware and software solution for robust imaging. Using this platform and CAD software, a 3D printable microscope was designed that is capable of birefringence microscopy, (Figure \ref{Figure 1.}) shows the design. After adoption, several modifications were made to the standard motorised OpenFlexure microscope. The key modifications were adapting the illumination dovetail (extending the length and increasing the strength of attachment), designing new parts to hold each of the polarised films, and accommodating ModLight light sources in the system \cite{modlight}. 
Building on this platform within the ModMicro suite of microscopy hardware, we designed Pol-ModMicro. The versatile \textbf{pol}arised light \textbf{mod}ular \textbf{micro}scope includes:
\\
\begin{enumerate}[nolistsep]
    \item an updated illumination dovetail, 
    \item a modified  condenser arm, with updated elements to hold the light pipe and incorporate ModLight \cite{modlight},
    \item a component for a rotating polariser, and
    \item a modified cube between the objective and the camera for the static second polariser.
    \\
\end{enumerate}

\begin{figure}[t]
\centering
\captionsetup{width=\linewidth}
\includegraphics[width=0.85\columnwidth]{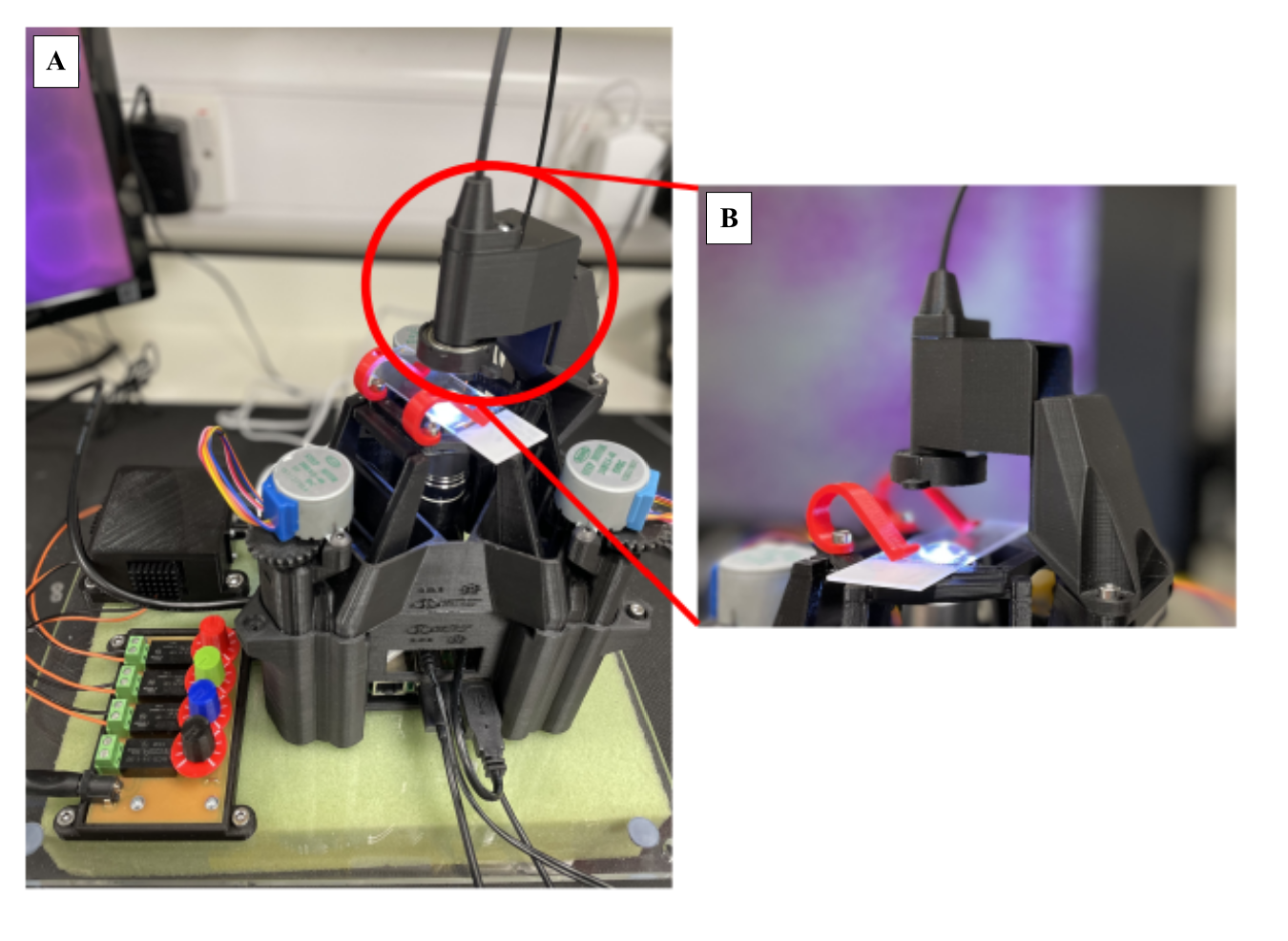}
\caption{\label{Figure 2.} Pol-ModMicro is an inexpensive, versatile and portable system for polarised light microscopy, as seen in (A). The inset (B) shows the modified condenser arm (with the light pipe and the rotating polariser) and the reinforced illumination dovetail used to establish cross-polarisation. The acrylic platform and foam sheet are used to enhance portability and negate vibrations.}
\end{figure}

Pol-ModMicro (Figure \ref{Figure 1.}) includes the modular X-Cube version of ModLight \cite{modlight}, allowing control of the spectral band and brightness of the illumination light. The sample in the microscope is illuminated using a light pipe, which is placed in a bespoke holder in the modified condenser arm (Figure \ref{Figure 2.}). This modified condenser also allows a rotating polariser (cut from LPVISE2X2, Thorlabs, Inc.) to be placed in the light path before it interacts with the sample. The polariser film is cut to 17 by 17 mm size and fixed in this holder, denoted as `Rotating Polariser' in the figure~\ref{Figure 1.}. Finally, a modified cube is placed between the sample and Raspberry Pi camera housing the `Fixed Polariser', from the same film cut to 11 by 13 mm. This novel adoption of the OpenFlexure Delta Stage microscope was designed to compensate for the need to purchase an expensive microscope to view \textit{Plasmodium}-infected samples with cross-polarised light. CAD models of the modified microscope and illumination components are available with this article, or on request. 

Once the system is assembled, the OpenFlexure Connect packages are updated and installed on the Raspberry Pi. Using the OpenFlexure Connect application, a Z-stack of images is taken both with polarised and bright-field illumination to identify and compare the \textit{Plasmodium}-infected cells containing hemozoin crystals. Further, the data is compared to a standard bright-field microscope system for validation and characterisation of the system. 



\subsection{Imaging Birefringence} 
To image birefringence from the hemozoin crystals, the fixed polariser cube is inserted below the objective lens, meaning that there is now a polarised film both above and below the biological sample. The rotatable polarised film on the illumination module is then rotated using the handle on the holder until cross-polarisation is achieved. 

This practice can then be used on the birefringent images of \textit{Plasmodium}-infected cells as the difference in brightness is significantly higher when polarised light shines upon hemozoin. Each Z-stack of images is categorised using  machine learning techniques to allow a standard method of identifying the parasite within samples. In addition, ImageJ and machine learning techniques are used to set a brightness intensity threshold and then overlay the processed image on the brightfield image. Therefore, by setting a brightness pixel threshold the noise can be significantly reduced when viewing the parasites in stacks of images. 
The system's imaging protocol and validation include capturing data under (1) bright-field and (2) cross-polarised illumination conditions. Subsequently, (3) the scale and quality of the imaging are compared to a standard microscope system (Olympus BX43) as reference data. 

\begin{figure}[h!]
\centering
\captionsetup{width=\linewidth}
\includegraphics[width=\columnwidth]{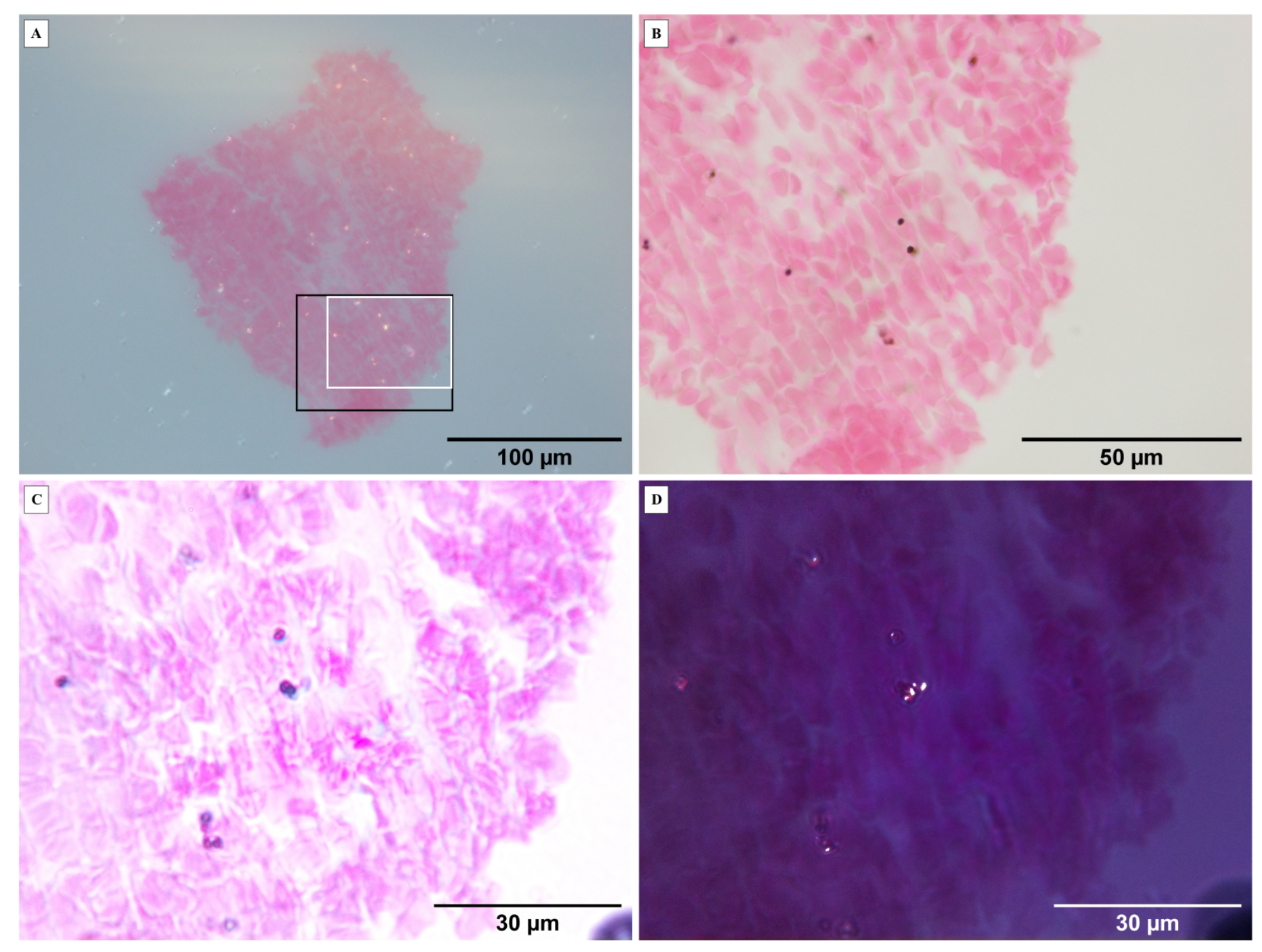}
\caption{\label{Figure 4.} This figure shows a combination of imaging techniques and systems at different resolutions. Using a 40x objective and the Olympus BX microscope, (A) shows the brightfield image of the cells in the blood smear with a single polariser in front of the LED light source. The black ROI indicates the imaging region of (B), while the white ROI indicates the same for (C) and (D). Using the same microscope, a second image is taken at 100x, and shown in (B). Finally, images (C) and (D) share brightfield and processed birefringent images captured by Pol-ModMicro with a 100x microscope objective.}
\end{figure}
The first image capture from Pol-ModMicro requires all the initial calibration systems that are considered standard for the OpenFlexure microscope. These images are collected without the rotating polariser, i.e. bright-field images. The rotating polariser is subsequently fixed in place and manually adjusted to achieve cross-polarisation, which in this case is assessed when the `darkest' image is captured. In this state, the birefringence from the hemozoin crystals is imaged. Image acquisition is performed by moving in a 2-micron step size in the Z direction, i.e. depth of the sample, to maximise the imaging efficiency while accounting for different orientations of the crystals in the sample itself. Qualitatively choosing the best-focused image, we set a brightness intensity threshold and overlay the processed image on the brightfield image. By setting a brightness pixel threshold, the noise can be significantly reduced when viewing the parasites in stacks of images.



\section{Results and Discussions}
The methodology illustrates brightfield and cross-polarised light imaging using a 3D-printed setup that can reliably acquire images. To compare the results with a standard microscope and simple polariser (as detailed in \cite{RAMADHANI}), we compare our imaging results with those obtained using an Olympus BX microscope. It is important to note that regions of interest, field of view and other imaging variables have an impact in achieving this comparison, and must also be given due consideration when applying this method. However, in the absence of a proprietary microscope, Pol-ModMicro has the advantages of being low-cost, robust, modular and portable.

In addition to using a high magnification objective, Pol-ModMicro captures images under polarised light with a good degree of accuracy, stability and repeatability (the position of the polariser can be marked/fixed to ensure cross-polarisation is set). The imaging system combined with the post-processing enables overlaying both datasets to accurately quantify the distribution of the parasites in the blood smear. In comparison to the 40x objective on the Olympus microscope, the hemozoin crystals are also substantially more identifiable. 

In this study, we acquire multiple stacks of high-resolution brightfield and birefringent images using a 3D-printed microscope and illumination system. These images are comparable to those captured with a more expensive, laboratory standard light microscope with a single polariser. This highlights the accessibility of the Pol-ModMicro system for low-cost birefringence imaging. Integration of ModLight source with the microscope is a key addition as it improves the illumination efficiency and imaging in ambient light and dark conditions. Adopting the Delta Stage  for Pol-ModMicro improves the sensitivity and step size resolution (Z-stack) to 0.1 $\mu$m \cite{McDermott}. The adapted parts of the microscope are now available, along with the OpenFlexure files (that remain unchanged), along with a step-by-step guide \cite{Eadie2022}. 


In light of these advantages, a few challenges must also be mentioned. These include:
\\
\begin{enumerate}[nolistsep]
    \item The three motors can cause slight drift when moving the stage of the microscope in large step sizes (approximately greater than 100 microns). This can be corrected through refocusing but creates some user difficulty when attempting to identify and image specific sites.
    \item The spherical aberrations are caused by the stage clips being screwed in as part of the design impacting the slide being clipped exactly horizontally.
    \item In the future a new design of the clips and the stage is necessary to reduce the negative impact of spherical aberration.
\end{enumerate}

\section{Conclusions}
Pol-ModMicro is a low-cost, multi-component and modular solution for brightfield and polarised light microscopy. The custom parts allow easy adaptability when changing between illumination strategies, and allow robust imaging of hemozoin crystals in \textit{Plasmodium} spp. Beyond this application, we also envisage this system being useful in applications where birefringence is a crucial differentiator. 


\begin{backmatter}
\bmsection{Funding}
This work was made possible with support from the Leverhulme Trust Early Career Fellowship, EPSRC Impact Acceleration Account (IAA) [EP/R511705/1] and EPSRC funding to QuantIC [EP/M01326X/1].

\bmsection{Acknowledgments}
The authors acknowledge the advice and support of Dr Richard Bowman (University of Glasgow), Mr Mark Main (University of Glasgow), Mr Robert Archibald (University of Glasgow) and Ms Olga Garcia (University of Glasgow) through the conception, execution and output of this work. 

\bmsection{Disclosures}
The authors declare no conflicts of interest.

\bmsection{Data availability} Data underlying the results presented in this paper are available in Ref. \cite{Eadie2022}.

\end{backmatter}
\bibliography{sample}
\end{document}